\documentclass[aps,pra,onecolumn,preprintnumbers,superscriptaddress]{revtex4-1}
\pdfoutput=1
\usepackage{amsmath}
\usepackage{bbold}
\usepackage{graphicx}
\usepackage{color}
\usepackage[T1]{fontenc}
\begin{document}
\title{Transport Enhancement of Irregular Optical Lattices with Polychromatic Amplitude Modulation}
\author{R. A. Pepino}
\author{W. P. Teh}
\affiliation{Florida Southern College, Department of Physics and Chemistry, Lakeland, FL 33801}
\author{L. J. Magness}
\affiliation{Rhodes College, Department of Physics, Memphis, TN 38112}
\date{\today}

\begin{abstract}
We demonstrate that the transport characteristics of deep optical lattices with one or multiple off-resonant external energy offsets can be greatly-enhanced by modulating the lattice depth in an exotic way.  We derive effective stationary models for our proposed modulation schemes in the strongly interacting limit, where only one particle can occupy any given site.  Afterwards we discuss the modifications necessary to recover transport when more than one particle may occupy the lattice sites.    For the specific five-site lattices discussed, we numerically predict transport gains for ranging from $4.7\times 10^6$ to $9.8\times 10^{8}$.
\end{abstract}

\maketitle
\section{Introduction}
Ultracold atoms in optical lattice potentials are very rich and diverse physical systems.  Their coherent dynamics make them candidates for quantum logic gates~\cite{Deutsch,brennen1999entangling}.  It has also been proposed that atomtronics components~\cite{myprl, eckel2014hysteresis,benseny2010atomtronics} can be realized in Bose-Hubbard lattices.  The typical scheme that might generate such components would involve customized lattices that exhibit certain transport characteristics.  Holographic mask techniques~\cite{bakr2009quantum} make it possible to realize such systems in optical lattices, while the current ability to image single atoms makes signal detection feasible.  Optical lattices have also proven to be excellent quantum simulators: not only has the mott-insulator to superfluid phase transition been realized~\cite{GreinerBloch}, but {\em lattice shaking}~\cite{lin2009synthetic,jaksch2003creation}, a modulation technique that involves periodically displacing the location of the wells, has recently been used to realize the Haldane model~\cite{esslinger} as well as the Meissner effect~\cite{atala2014observation} in these systems.  This lattice shaking technique has also been used to realize {\em photon-assisted tunneling}~\cite{sias2008observation,weiss2009photon,kolovsky2011creating}, in which the modulation is used to facilitate transport in otherwise frustrated optical lattice systems.

An alternative modulation technique known as {\em amplitude modulation} involves varying the depth of the optical lattice $V(t)$ sinusoidally.  This method, referred to as AC-shaking, is also capable of creating synthetic gauge fields in optical lattice systems~\cite{hauke2012non}.  This technique was originally developed as a spectroscopic tool to study optical lattices~\cite{stoeferle,kollath}.  It was then proposed as a means to populate higher Bloch bands of the optical lattice~\cite{sowinski2012creation,lacki2,lacki}.    More recently, multiple frequency or {\em polychromatic}  modulations were used to stimulate two-photon transitions that coherently promoted a significant number of atoms from the ground to the fourth excited Bloch band~\cite{niu}.  Photon-assisted tunneling has also been achieved in systems using amplitude modulation~\cite{greinerPAT,chen2011controlling}.

This article is concerned with optimizing the transport characteristics of irregular optical lattices with  external energy offsets that would otherwise greatly reduce the transport within the lattice.  The work presented can be viewed as a generalization of photon assisted tunneling with amplitude modulation.  Here we study optical lattices where $V(t)$ is not necessarily varied perturbatively or sinusoidally, and there are potentially several unique energy offsets within the lattice structure. 
\begin{figure}[h]
\includegraphics[width=.4\textwidth]{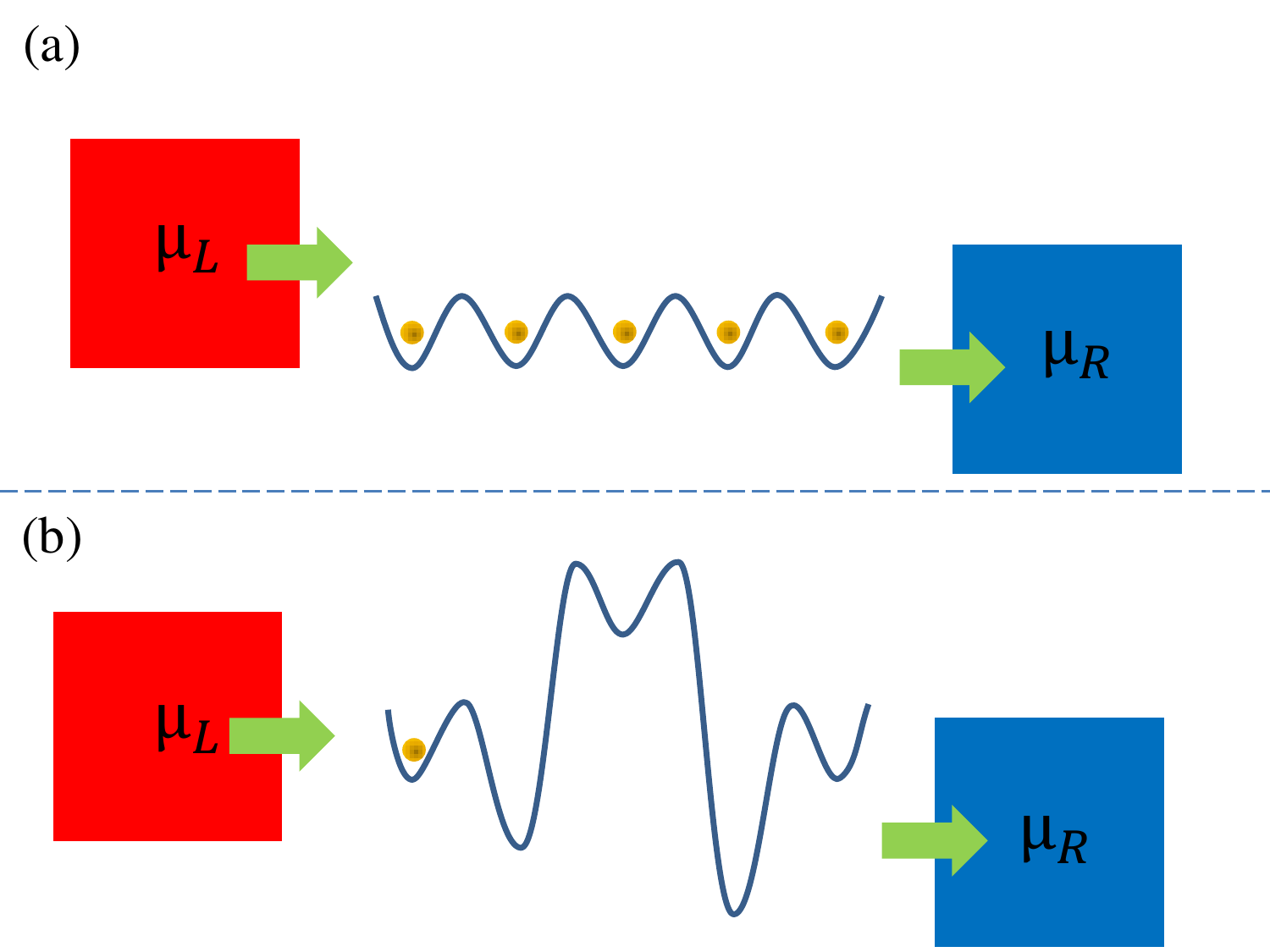}
\caption{An illustration of a one-dimensional optical lattice whose end sites are driven by ultracold atomic reservoirs. In the strongly-interacting case where the reservoirs are conditioned to put only one atom into the system, the energetically flat lattice depicted in (a) yields an optimal transport  response, whereas the lattice depicted in (b) yields virtually zero transport.}\label{toon}
\end{figure}
The immediate systems that we have in mind are customized optical lattices, produced possibly by holographic masks and/or other techniques that can generate such lattices~\cite{Ma}, that are designed to have exotic transport characteristics.  For instance, the development of synchronous atomtronics logic elements, may require transport across coupled lattice sites with energy mismatches that are large compared to the tunneling rates.  However, the results we present are very general, and they can be applied to a variety of systems:  although we analyze the one-dimensional open bosonic case in this work, a variation of the presented schemes are valid for either bosons or fermions, trapped in open or closed quantum systems comprised in one-, two-, or three-dimensional optical lattices or possibly atom-chip experiments.  They might find applications in an experiment where the control of a coherent atomic signal is desired.   Finally, the modulation schemes might be useful as a spectroscopic tool: scanning through modulation frequencies of an arbitrary lattice may lead to spikes in the transport response, which could be used to infer properties of the structure of the lattice itself.  Such a tool may be useful for studies of disordered optical lattices~\cite{damski2003atomic,schulte2005routes,pasienski2010disordered}.

We demonstrate that the transport response of irregular, deep, lowest band lattices can be significantly enhanced by a polychromatic modulation of the {\em tunneling rates} of the system.  As discussed below, this requires a unique periodic modulation of the lattice depth itself. In a regime where the external energy differences are large compared to the intrasystem tunneling, we demonstrate that this modulation enhances transport by mapping the formerly stationary irregular system Hamiltonian that prohibits transport onto an effective {\em stationary} Hamiltonian that accurately predicts the optimized transport response.  

We examine the transport properties of these systems by driving their end sites with reservoirs of neutral, ultracold atoms. When a chemical potential difference exists between the reservoirs, atomic transport (or current) may be induced across the system.  For example, when the reservoirs and the lattice are conditioned so that each system site can support at most one atom, the flat energy landscape depicted in Fig.~\ref{toon}(a) allows the atoms to freely explore the entire lattice.  This configuration yields the optimal transport response.  If the landscape has one or more external energy differences, as in Fig.~\ref{toon}(b), the current can be greatly suppressed.  The schemes presented in this work map a system such as the one illustrated in Fig.~\ref{toon}(b) onto an effective lattice of the type in Fig.~\ref{toon}(a).

The conditions for transport optimization turn out to depend on the maximum site occupancy, which is determined by the value of the driving chemical potential.  After examining the case where a maximum of one atom may exist on any given site, we follow with a discussion of current recovery for situations involving  higher site occupancies.

The structure of this article goes as follows: in Sec.~\ref{th} we introduce the time-dependent system Hamiltonian, discuss our open quantum system theory and define the observable that we refer to as `transport current'.  In Sec.~\ref{sbs} we analyze the single site occupancy case.  A unitary transformation is used to demonstrate the formal equivalence between stationary flat lattices and arbitrary lattices whose tunneling rates undergo a specific complex phase modulation.  We then adapt this reasoning to propose an amplitude modulation scheme that leads to large gains in an irregular lattice's transport response.  Employment of a secular approximation allows us to derive effective stationary models that accurately reproduce the current dynamics of these complicated offset modulated systems.  Afterwards, numerical simulations are provided to confirm and characterize the recovered transport response.  In Sec.~\ref{mae} we discuss transport recovery for the case where up to two atoms can occupy any given lattice site.  Extensions to higher site occupancies is straight-forward. 

\section{Theory}\label{th}
\subsection{System Hamiltonian}
The systems we consider in this article are ultracold bosonic atoms trapped in an irregular one-dimensional optical lattices whose lattice depth $V(t)$ can be modulated in time.  The lattice potential is assumed to be 
$V_{lat}(x,t)=V(t)\sin^2(\pi x/a_l)+V_{ext}(x)$,
where $a_l$ is the lattice spacing and $V_{ext}$, such that $|V_{ext}(x)|\ll V(t)$ for all $x$ and $t$, is the external potential that generates the energy shifts in the lattice structure.

Working in the lowest Bloch band and in the tight-binding regime, the stationary system dynamics are well-described by the Bose-Hubbard model~\cite{zoller}.  If the modulation timescale is slow compared to the system dynamics, then the time evolution of the quantum states can be treated adiabatically~\cite{lacki}.  Under these assumptions, we can model the Bose-Hubbard system as a function of the lattice depth $V$ as
\begin{equation}\label{gbh}
\hat{H}_S(V)=\sum_{j=1}^N \omega_j(V)~\hat{a}^\dagger_j\hat{a}_j+\frac{1}{2}\sum_{j=1}^NU_j(V)\ \hat{a}^\dagger_j\hat{a}_j^\dagger\hat{a}_j\hat{a}_j
-\sum_{j=2}^{N}\left(J_j(V)~\hat{a}_{j}^\dagger\hat{a}_{j-1}+J_j^*(V)~\hat{a}_{j-1}^\dagger\hat{a}_j\right) \ ,
\end{equation}
where $\omega_j$ is the external site energy on $j$, $\hat{a}^\dagger_j$ ($\hat{a}_j$) creates (annihilates) a particle on site $j$, $U_j$ is the on-site interaction energy and $J_j$ is the nearest neighbor tunneling matrix element between sites $j-1$ and $j$.

The deep lattice assumption allows us to approximate the dependence of the system parameters on $V$ by treating the wells as coupled harmonic oscillators~\cite{zwerger2003mott}.  Under these assumptions, and in a frame rotating with the ground state energy of the perpendicular trapping potential, the Bose-Hubbard parameters have the following dependence on $V$:
\begin{eqnarray}
\omega_j(V)&=&\int_{-\infty}^\infty dx \ \psi_j^*(x)\left(\frac{\hat{p}^2}{2m}+\hat{V}_{lat}(x)\right)\psi_j(x)\nonumber\\
U_j(V)&=&\frac{4\pi a_s\hbar^2}{m}\int  d^3\vec{r} \ |\psi_j(x)\psi_\perp(y,z)|^4\\
J_j(V)&=&\int_{-\infty}^\infty dx \ \psi^*_j(x) \left(\frac{\hat{p}^2}{2m}+V_{lat}(x)\right)\psi_{j-1}(x)\nonumber
\end{eqnarray}
where $\psi_j(x)$ is the ground state harmonic oscillator solution of the local lattice potential centered at site $j$, $a_s$ and $m$ are the scattering length and mass of the atom, and $\psi_\perp(y,z)$ is the harmonic oscillator solution of the trapping potential.  Analytic solutions for $\omega_j(V)$ and $U_j(V)$ are 
\begin{eqnarray}
\omega_j(V)=2\sqrt{E_r}\sqrt{V}+V_{ext}(x_j)\nonumber\\
U_j(V)=\sqrt{8\pi}\frac{a_s}{a_l}E_r^{1/4}\sqrt{V_\perp}V^{1/4}
\end{eqnarray}
where all of the parameters are expressed in terms of recoil Energies $E_r=\hbar^2\pi^2/(2ma_l^2)$. Throughout this work we assume a lattice spacing of $a_l=640$~nm, the mass for $^{87}\text{Rb}$ whose scattering length is taken to be $a_s=5.2$~nm, and the perpendicular lattice depth is ${V_\perp=50 E_r}$.  In our model, we obtain $J_j(V)$ numerically.

Previous works proposed by Refs.~\cite{sowinski2012creation,lacki} have demonstrated that sinusoidal modulation of $V(t)$ can lead to higher Bloch band excitations.  However, assuming that $V(t)\sim 15 E_r$ guarantees a separation of parameters ${\omega\gg U\gg J}$.  Since the modulations presented below will consist of frequencies $\alpha\leq U$, higher Bloch band dynamics can be safely neglected.  This has been verified in our numerical simulations of higher band dynamics these systems.  

\subsection{Time-Dependent Born-Markov Master Equation}
We study the transport properties of these one-dimensional systems using the tools developed in \cite{mypra}, where the ends of the lattices are coupled to reservoirs acting as sources and sinks of ultracold atoms as in Fig.~\ref{toon}.  We use a quantum master equation approach \cite{zubarev,cohen} to model the dynamics of the Bose-Hubbard system coupled to reservoirs of ultracold neutral atoms.  Working in the zero-temperature limit with strongly repulsive atoms, the reservoir's level occupancy can be characterized by a chemical potential $\mu$, such that all states below $\mu$ are occupied and all states above are vacant.  A chemical potential difference between the left and right reservoirs $\mu_L>\mu_R$, may induce an atomic current across the lattice from left to right.   When $U\gg J$ the particle manifolds of the system become well-separated.  If the values of $\mu_{L,R}$ do not resonantly overlap with any system eigenenergies, the care taken in Ref.~\cite{mypra} is not necessary, and we can employ the standard Born and Markov approximations \cite{cohen,walls,meystre}.  Under these conditions, the value of $\mu_L$ determines the maximum allowed particles on any given site.  In our model, we assert the value of $\mu_L$ by truncating the system's basis at the appropriate level.

Since the modulation frequencies $\alpha$ will be of the same order of magnitude as the other frequencies associated with the system, the rapid decay of the memory kernel ensures that the time-dependence of the system has a negligible affect on the Liouvilian.  This generates a master equation of the standard Lindblad form \cite{mythesis}.  For an N-site lattice where atoms are being pumped onto site $1$ by a reservoir with chemical potential $\mu_L$, and removed from site $N$ by a reservoir with chemical potential $\mu_R=0$, our master equation model is
\begin{equation}\label{meq}
\frac{d\hat{\rho}}{dt}=-i\left[\hat{H}_S(t),\hat{\rho}\right]
-\frac{\kappa}{2}\left(\hat{a}_1\hat{a}_1^\dagger\hat{\rho}+\hat{\rho}\hat{a}_1\hat{a}_1^\dagger-2\hat{a}_1^\dagger\hat{\rho}\hat{a}_1\right)
-\frac{\kappa}{2}\left(\hat{a}_N^\dagger\hat{a}_N\hat{\rho}+\hat{\rho}\hat{a}^\dagger_N\hat{a}_N-2\hat{a}_N\hat{\rho}\hat{a}^\dagger_N\right),
\end{equation}
where $H_S(t)=H_S(V(t))$, and $\kappa$ is the decay rate of the reservoirs, which we assume to be identical for convenience.

\subsection{Transport Current}
\begin{figure}[h]
\includegraphics[width=.37\textwidth]{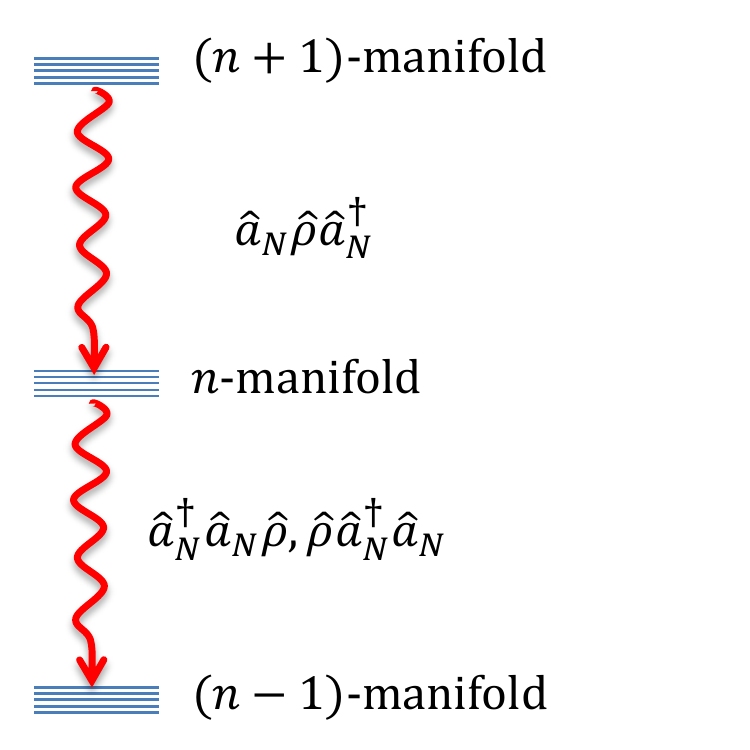}
\caption{An illustration of the actions of the reservoir connected to the $N^{\rm th}$ lattice site on an $n$-particle  system state.  The $\hat{a}_N\hat{\rho}\hat{a}_N^\dagger$ term contributes to population of the $n$-particle state from the $(n+1)$ particle manifold, while the $\hat{a}_N^\dagger\hat{a}_N\hat{\rho}$ and $\hat{\rho}\hat{a}_N^\dagger \hat{a}_N$ terms deplete population to the $(n-1)$ particle manifold.}\label{res}
\end{figure}

We define the atomic transport, or current, in these systems by monitoring the average number of particles leaving the system.  This can be determined from the rate contributions of the master equation.  The reservoir coupled to the $N^{\rm th}$ lattice site is responsible for particles leaving the system.  This reservoir action is quantified in the last line of Eq.\eqref{meq}.  In Fig.~\ref{res}, we illustrate how the action of each of these Liouvillian terms affects the system populations.  Taking the inner product of Eq.~\eqref{meq} with a system eigenstate $|j\rangle$ who belongs to the $n$ particle manifold of the system (for example) shows that the $\kappa \langle j|\hat{a}_N\hat{\rho}\hat{a}_N^\dagger|j\rangle$ term corresponds to positive contributions, populating the state $|j\rangle$ from the $n+1$ particle manifold, while the $-\kappa/2\langle j|\hat{a}^\dagger_N\hat{a}_N\hat{\rho}|j\rangle$ and $-\kappa/2\langle j|\hat{a}^\dagger_N\hat{a}_N\hat{\rho}|j\rangle$ terms lead to depletion of state $|j\rangle$ to the $n-1$ particle manifold.  Therefore, the net current can be defined as either the total populating of the system from all states above, or the total depletion to all states below.  We choose the former, yielding:
\begin{equation}
\langle	\hat{I}\rangle=\sum_j  \kappa \langle j|\hat{a}_N\hat{\rho}\hat{a}_N^\dagger|j\rangle=\kappa {\rm Tr}[ \hat{a}_N\hat{\rho}\hat{a}_N^\dagger]=\kappa \langle\hat{a}_N^\dagger \hat{a}_N\rangle \ ,
\end{equation}
where we have exploited the cyclic property of the trace, and recognize 
\begin{equation}\label{current}
I=\kappa \langle  \hat{a}_N^\dagger\hat{a}_N \rangle
\end{equation}
as the current out of the system.  This definition makes sense, since $\kappa$ is the decay rate of the Lindblad master equation.  Thus the current out should equal the average number of particles on the $N^{\rm th}$ site, multiplied by the rate at which the particles leave the system.

\section{Single Atom Excitations}\label{sbs}
We first consider cases where $U\gg J$, and $\mu_L$ is set to maintain an occupancy of one atom on the first site.

\subsection{Modulating the Optical Lattice}\label{mod}
Consider a stationary ($V(t)=V_{\min}$) $N$-site system defined by Eq.~\eqref{gbh} with arbitrary external energies $\omega_j$.  When energy differences between at least one pair of adjacent sites  are large compared to their tunneling amplitudes, transport is greatly suppressed. Rotating this system with
\begin{equation}\label{rot}
\hat{H}_0=\sum_{j=1}^N\omega_j\hat{a}_j^\dagger\hat{a}_j
\end{equation}
yields
\begin{equation}\label{rbh}
\tilde{H}(t)=\frac{1}{2}\sum_{j=1}^N U_j\hat{a}_j^\dagger\hat{a}_j^\dagger\hat{a}_j\hat{a}_j
-\sum_{j=2}^{N}\left(J_j ~ e^{i\delta_j t} ~ \hat{a}_{j}^\dagger\hat{a}_{j-1}+J_j^*~e^{-i\delta_j t} ~ \hat{a}_{j-1}^\dagger\hat{a}_j\right) \ ,
\end{equation}
where $\delta_j=\omega_j-\omega_{j-1}$, is the energy difference between sites $j$ and $j-1$.  

If each tunneling rate in Eq.~(\ref{rbh}) could be modulated with
\begin{equation}\label{jideal}
J_j \to J_j e^{-i \delta_j t} \ ,
\end{equation}
then the offset modulated system of Eq.~(\ref{rbh}) is equivalent to a flat stationary lattice system, which would yeild the optimal current response.  The modulation  in Eq.~(\ref{jideal}) is ideal, since it would fully recover the maximal transport response from an arbitrary offset stationary lattice.  The lesson learned from this example is that sinusoidal modulations of the tunneling rates may greatly enhance transport in lattices with energetic mismatches.


\subsection{Optimization Scheme}\label{optsch}
Given an arbitrary $N$-site lattice, we propose that if there are  ${M\leq N}$ unique, nonzero values of $|\delta_j|$, that a great deal of the intrasystem current can be recovered if the overall lattice depth is varied such that the tunneling barriers experience a sinusoidal modulation consisting of a superposition of the $M$ unique frequencies.  

For example,
\begin{equation}\label{globV}
V(t)=V_{\min}-\frac{1}{\beta}\ln\left[\frac{1}{M} \sum_{k=0}^M\cos^2\left(\frac{\delta_k}{2}t\right)\right]
\end{equation}
would yield
\begin{equation}\label{globH}
\hat{H}_S(t)\simeq\sum_{j=1}^N \omega_j(t)~\hat{a}^\dagger_j\hat{a}_j+\frac{1}{2}\sum_{j=1}^NU_j(t)\ \hat{a}^\dagger_j\hat{a}_j^\dagger\hat{a}_j\hat{a}_j
-\frac{1}{M}\sum_{k=1}^{M} \cos^2\left(\frac{\delta_k}{2}t\right)\sum_{j=2}^NJ_j\hat{a}_{j}^\dagger\hat{a}_{j-1}+\text{H.c.} \ .
\end{equation}
Here, $\beta$ is the parameter from an exponential fit of the tunneling $J(V)\simeq  J_{\rm max} e^{-\beta (V-V_{\rm min})}$. Assuming $V_{\rm min}= 15 E_r$, we find $J_{\rm max}\simeq 6\times 10^{-3} E_r$ and $\beta\simeq 0.24/E_r$.  In order to compare results against the response of a stationary ideal lattice with a tunneling rate of $J_{\rm max}$, we've included the $1/M$ term to guarantee that the tunneling rates in the modulated Hamiltonian do not exceed $J_{\rm max}$.

Although a clever choice of the relative phases of the oscillatory drive can enhance the current response even further, we adopt cosines in this work to yield a simpler analytic form.

\subsection{Effective Stationary Models}\label{EffSec}
In the deep lattice regime, {\em stationary} effective flat lattice models can be generated that accurately reproduce the dynamics of Eq~\eqref{globH}.  These effective models are significantly less demanding to numerically evolve.  They also shed light on the transport recovery mechanism proposed above.

Assuming the global modulation from Eq.~(\ref{globV}), expressing the onsite interaction energy as a modulation about its mean $U(t)=\langle U \rangle+\Delta U(t)$ and transforming into an interaction picture that rotates with the time-dependent version of Eq.~(\ref{rot}) yields
\begin{equation}\label{gwigs}
\begin{split}
\tilde{H}_S(t)=&
\frac{1}{2}\sum_{j=1}^N \left(\langle U_j\rangle+\Delta U_j(t)\right)\ \hat{a}^\dagger_j\hat{a}_j^\dagger\hat{a}_j\hat{a}_j\\
&-\left(\frac{1}{2}+\frac{1}{4 M}\sum_{k=1}^{M} e^{i\delta_k t}+e^{-i\delta_k t}
\right) 
\times\sum_{j=2}^NJ_j(t)e^{i\delta_j t}~\hat{a}_{j}^\dagger\hat{a}_{j-1}
+\text{H.c.}
\end{split}
\end{equation}
where $\delta_j(t)\simeq\delta_j$ in the assumed parameter regime.

When two adjacent sites have the same energy, the stationary portion of the tunneling matrix element between these sites is $J/2$.  When $\delta_j\neq 0$, the stationary portion becomes $J/(4M)$.  Since $|\delta_j|\gg \Delta U,J$ a secular approximation similar to the rotating wave approximation can be made: for timescales large compared to ${\hbar/\min\{|\delta_j|\}}$, the oscillating terms in Eq.~(\ref{gwigs}) become insignificant compared to their stationary counterparts.  Neglecting these terms yields
\begin{equation}\label{StEff}
\hat{H}_\text{eff}=
\frac{1}{2}\sum_{j=1}^N \langle U_j\rangle \hat{a}^\dagger_j\hat{a}_j^\dagger\hat{a}_j\hat{a}_j
-\sum_{j=2}^N \tilde{J}_j~\hat{a}_{j}^\dagger\hat{a}_{j-1}
+\text{H.c.}
\end{equation}
where
\begin{equation}\begin{split}\label{sbsRules}
\delta_j=0\quad & \implies\quad \tilde{J}_j=\frac{J}{2}\\
\delta_j  \neq 0\quad & \implies \quad \tilde{J}_j=\frac{J}{4 M} \ .
\end{split}
\end{equation}
That is, the assumed modulation of the gapped system yields an effective stationary Hamiltonian whose flat energy landscape assures the optimal current response.  Although the rescaling of the tunneling rates implies a current response less than the corresponding flat lattice with uniform $J$, the current gained in these systems is still several orders of magnitude greater than the unmodulated lattice with one or more energy offsets.

\subsection{Numerical Results}
\begin{figure}[h!]
\includegraphics[width=.6\textwidth]{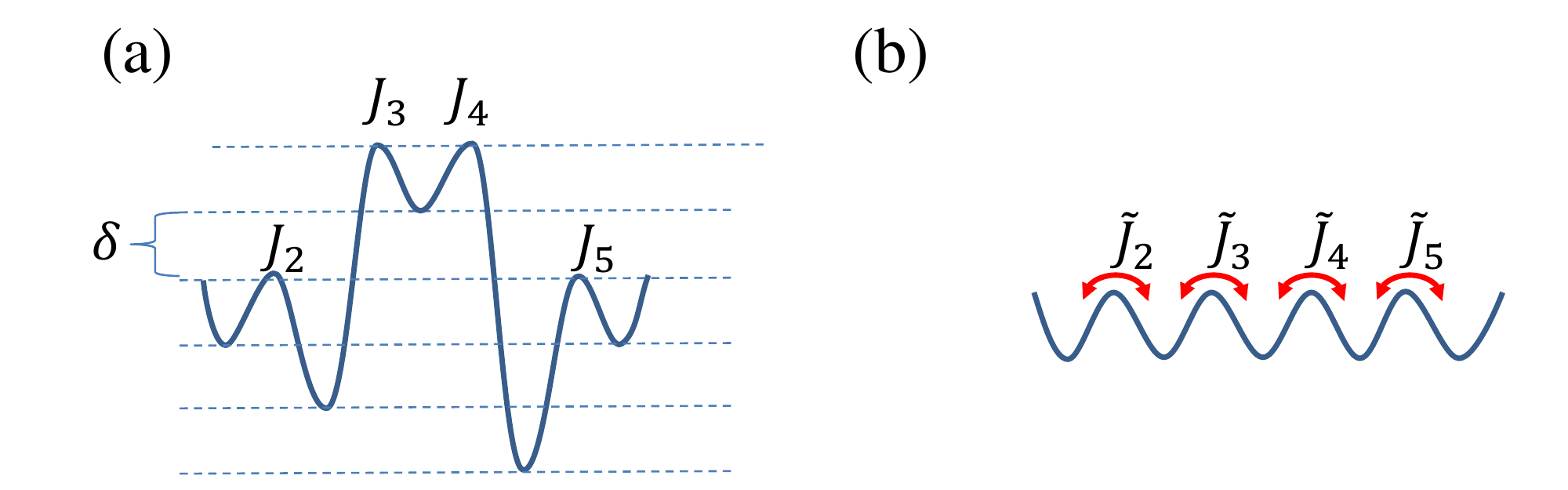}
\caption{The mapping of the five-site optical lattice depicted in Fig.~\ref{toon} (a) onto its effective stationary Hamilton when modulated (b).  In this example $\tilde{J}_j=J_j/16$.}\label{FiveEff}
\end{figure}

In this section, we numerically verify transport recovery of Eq.~\eqref{gbh} due to the modulation scheme presented in Sec.~\ref{optsch}, and the accuracy of the effective stationary model in Sec.~\ref{EffSec}, for a variety of five-site lattices.  We first consider the system depicted in Fig.~\ref{FiveEff}(a).  Working with a minimum lattice depth of $V_{\rm min}=15 E_r$, we assume ${\delta_j=[-0.1,0.3,-0.4,0.2] E_r}$.  This lattice has four unique energy offsets.  When coupled to reservoirs whose decay rates satisfy $J_j/\kappa\simeq 15$, this lattice exhibits a steady-state current response that is reduced by a factor of $1.3\times 10^9$ compared to the maximum response determined by its flat counterpart.  Since we are assuming that $\mu_L$ is set to maintain an occupancy of one particle on the left site in this section, the values $U_j$ are irrelevant since the onsite interaction terms are not accessed.

\begin{figure}[h!]
\includegraphics[width=.48\textwidth]{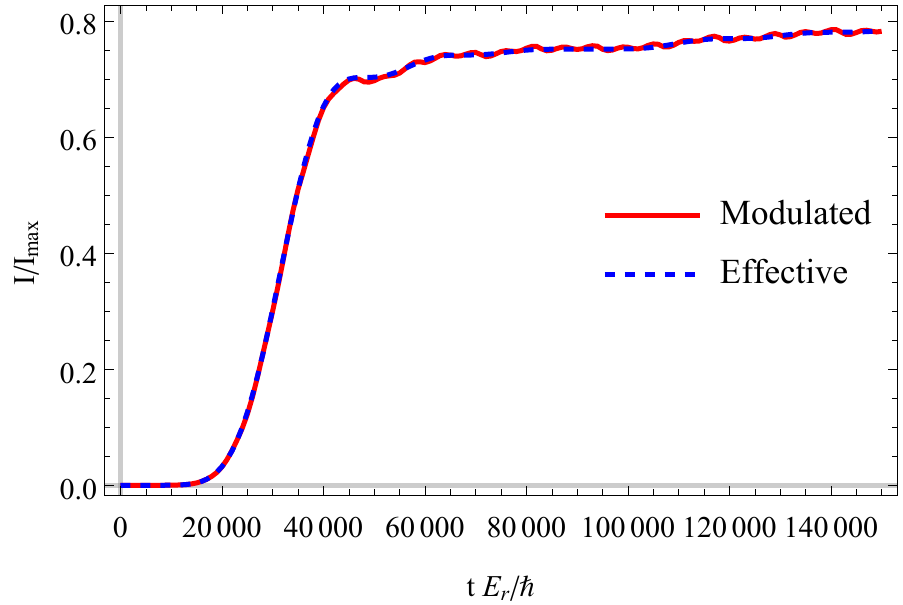}
\caption{Time evolution of the current response of the modulated, offset lattice depicted in Fig.~\ref{FiveEff}(a) (red solid) and the stationary effective Hamiltonian in Fig.~\ref{FiveEff}(b) (blue dashed). The modulation increases the lattice's current response by a factor greater than $9.8\times 10^{8}$, recovering over $60\%$ of the lattice's maximal current response.  The stationary effective model accurately captures the evolution for times large compared to $t=\hbar/E_r$.}
\label{FiveEvol}
\end{figure}

The effective stationary model for the dynamics of this lattice under the proposed modulation is illustrated in Fig.~\ref{FiveEff}(b).
A numerical simulation of the full unapproximated modulated dynamics of Eq.~\eqref{gwigs} for this energy-offset lattice is provided in Fig.~\ref{FiveEvol}.  The current response is normalized by the steady-state maximal response of the flat, stationary lattice with uniform $J$.  In steady-state, modulation leads to a current gain of $9.8\times 10^{8}$ compared to the stationary offset lattice, which is $78\%$ of the current response of the ideal system.  The effective stationary model of Eq~\eqref{StEff} is also plotted in this figure.  For times large compared to timescales set by the energy differences of this system ($t\sim\hbar/E_r$ in the plot), the unique dynamics of the modulated system are accurately recovered by the effective model, yielding a percent error of $0.2\%$ from the full steady-state dynamics.

The steady-state current gains and accuracy of the corresponding effective models for a variety of five-site lattices are provided in Table~\ref{Table}.  The characteristics of the Hamiltonian described above are presented in the fourth row of this table.
\begin{table}[h]
\begin{tabular}{| c | c  | c|c |}
\hline
$\delta_j (E_r)$ 			& ~Current Gain~ 		& ~Percent  Recovered~		&~Percent Error of $H_{\rm eff}$ ~ \\ \hline  
$[0,0.1,0,0]$				&  $5.6\times 10^6$			&		64\%						& $1.0\%$\\ \hline
$[0.2,0,-0.2,0]$			&  $3.3\times 10^7$			&		46\%						& $1.5\%$\\ \hline
$[0,0.1,0.3,-0.3]$			&	 $4.7\times 10^6$		&		13\%						&	$0.1\%$\\ \hline
$[-0.1,0.3,-0.4,0.2]$		&  $9.8\times 10^{8}$		&		78\%						& $0.2\%$ \\ \hline
\end{tabular}
\caption{The current characteristics of a variety of five-site lattices.  The columns display the lattice characterized by its energy gaps, the steady-state current gain from modulation, the percent recovered compared to an ideal system and percent error of the corresponding effective Hamiltonian.}\label{Table}
\end{table}
In all four cases, the atomic current is largely recovered and the effective models accurately predict the dynamics of each modulated system.

As $V_{\rm min}$ gets reduced, the proposed modulation scheme is still capable of recovering most of the current.  For example, assuming that $V_{\rm min}=5Er$, $J/\kappa\simeq 29$ and that a Feshbach resonance can increase the scattering length so that the $U_j$ terms remain irrelevant, simulations show that the $\delta_j=[-0.1,0.3,-0.4,0.2] E_r$  lattice experiences a current gain of $\simeq 4.7\times 10^5$ where over $71\%$ of the maximum current is recovered.  The percent error of our effective model in this case is $30\%$.  The effective model is breaking down since we are starting to push the limits of our approximation, but the current gained in the system remains significant.

\subsection{Gain as a function of maximum lattice depth}
\begin{figure}[h!]
\includegraphics[width=.45\textwidth]{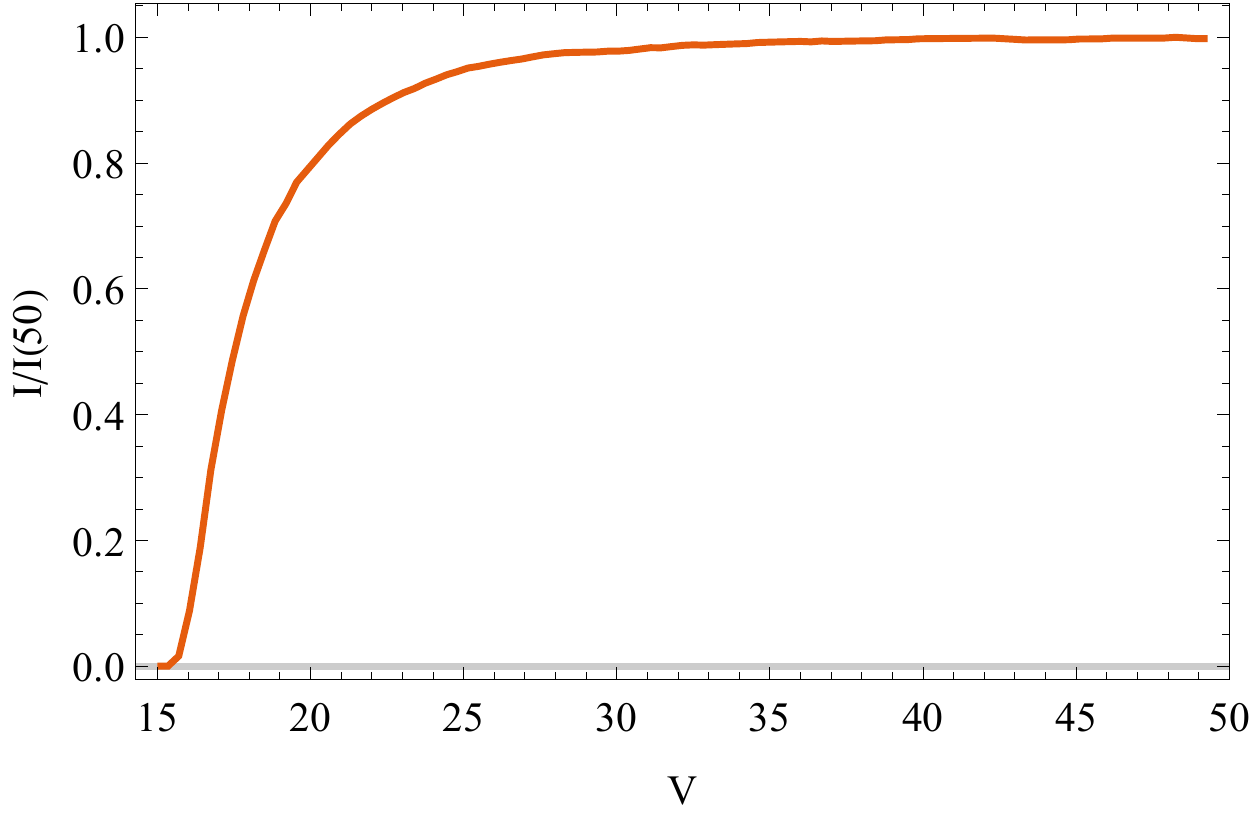}
\caption{Current response of the modulated lattice in Fig.~\ref{FiveEff}(a) as a function of $V_{\max}$ assuming $V_{\min}=15 E_r$.}\label{delJ}
\end{figure}

In an experiment, it might not be advantageous or possible to vary the tunneling from some maximal $J$ to zero. Next we examine the current response of the Hamiltonian depicted in Fig.~\ref{FiveEff}(a) as a function of the maximum lattice depth $V_{\max}$ assuming ${V_{\min}=15 E_r}$.  Figure~\ref{delJ} presents this simulation where the current response is normalized to its steady-state value when $V_{\max}=50 E_r$.
The data from this plot shows that varying the lattice depth by $2 E_r$, $5 E_r$, or $8 E_r$ recovers respectively $38\%$, $79\%$, or $91\%$ of the maximum current obtained when $V_{\max}=50 E_r$.  Thus, even small depth changes lead to significant current gains.

\section{Multiple Atom Excitations}\label{mae}
The optimization condition changes when the left reservoir is set to maintain an occupancy greater than one particle on the left site.  
We close by considering a two-site lattice with a finite energy offset $\delta_2$, where $\mu_L$ is set to maintain an occupancy of two atoms on the first site.  For an arbitrary modulation frequency $\alpha$, the system Hamiltonian we consider is
\begin{equation}
\begin{split}
\hat{H}(t)=~&
\omega_1(t)\hat{a}^\dagger_1\hat{a}_1
+\Big( \omega_1(t)+\delta_2(t)\Big)\hat{a}^\dagger_2\hat{a}_2\\
&+\frac{\langle U\rangle +\Delta U(t)}{2}\left(\hat{a}_1^\dagger\hat{a}^\dagger_1\hat{a}_1\hat{a}_1+
\hat{a}^\dagger_2\hat{a}^\dagger_2\hat{a}_2\hat{a}_2\right)\\
&-J \cos^2\left(\frac{\alpha}{2}t\right)\left(\hat{a}^\dagger_2\hat{a}_1+\hat{a}^\dagger_1\hat{a}_2\right) \ .
\end{split}\label{bh2}
\end{equation}
Assuming $V_{\min}=15 E_r$ and $V_{\max}=50 E_r$ implies $\delta_2(t)\simeq\delta_2$, $\langle U\rangle\gg|\Delta U(t)|$ and $\langle U\rangle/J\simeq 390$.  We futher assume $\delta_2=0.1 E_r$ and $J/\kappa\simeq 15$.

\begin{figure}[h!]
\includegraphics[width=.48\textwidth]{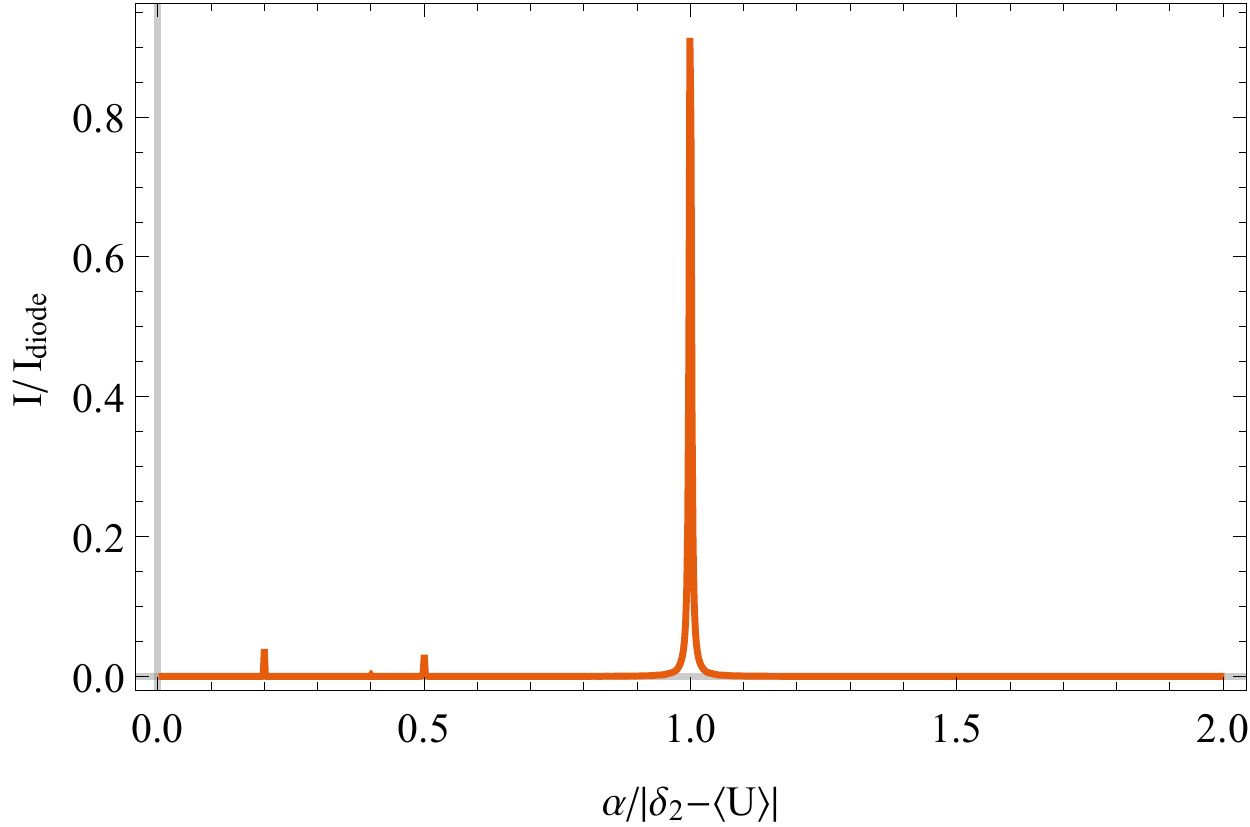}
\caption{Numerical current response for an energy mismatched two-site lattice driven to maintain an occupancy of two atoms on left site.  Modulating at $\delta_2$ (corresponding to $\simeq 0.2$ on the horizontal axis), leads to a current response that is small compared to the one obtained when modulating the lattice at $\delta_2-\langle U\rangle$.}\label{3stsw}
\end{figure}

The numerical steady-state current response of Eq.~\eqref{bh2} in Eq.~\eqref{meq} as a function of  the modulation frequency $\alpha$ is presented in Fig.~\ref{3stsw}.  
In the previous section, the current was greatly enhanced when the system was modulated at the frequency of the gap $\delta_2$. 
Here, when $\alpha=\delta_2$, the current increases by a factor of $7.3\times 10^3$.  However, when $\alpha=\delta_2-\langle U\rangle$, a current gain of $1.7 \times10^5$ is observed.   The effective stationary Hamiltonian corresponding to Eq.~(\ref{bh2}) with $\alpha=\delta_2-\langle U\rangle$ is 
\begin{equation}
\hat{H}_{\text{eff}}=~
\omega_1\hat{a}^\dagger_1\hat{a}_1
+(\omega_1+\langle U\rangle )\hat{a}^\dagger_2\hat{a}_2
+\frac{\langle U\rangle}{2}\left(\hat{a}_1^\dagger\hat{a}^\dagger_1\hat{a}_1\hat{a}_1+
\hat{a}^\dagger_2\hat{a}^\dagger_2\hat{a}_2\hat{a}_2\right)
-\frac{J}{4} \left(\hat{a}^\dagger_2\hat{a}_1+\hat{a}^\dagger_1\hat{a}_2\right) \ .
\label{bh2EffGapped}
\end{equation}
The percent error between current predicted by this effective model and the full modulated system is $7.2\%$.

\begin{figure}[h!]
\includegraphics[width=.45\textwidth]{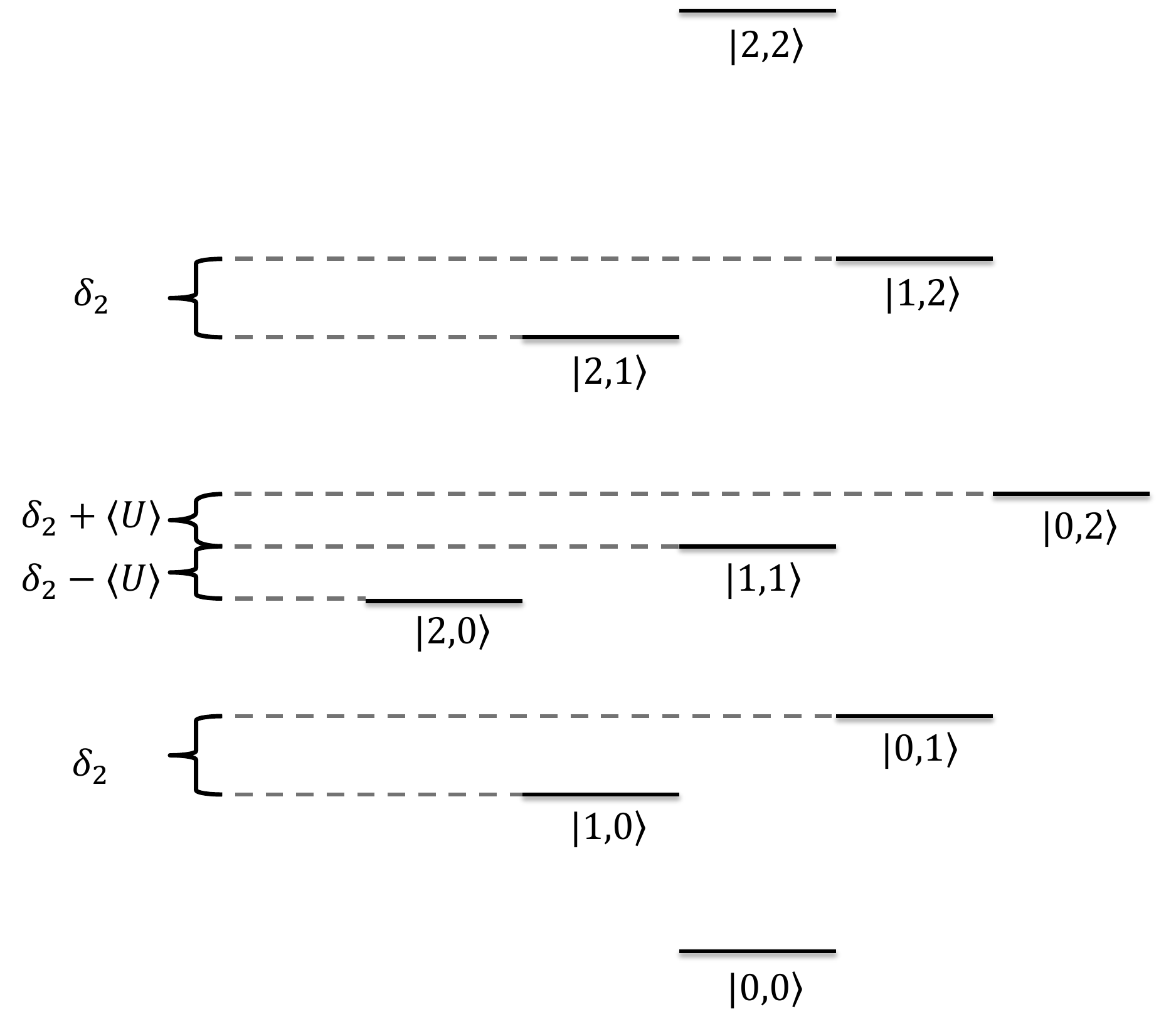}
\caption{An illustration of the relative relationship of the uncoupled Fock states, for zero to two particles per site, of a two-site lattice with finite $\delta_2$ and $\langle U\rangle$.}\label{baren}
\end{figure}

The change in the optimization condition can be understood by examining how the energetic relationships of the unmodulated system's Fock states effectively change when the system is modulated at frequencies $\delta_2$ and $\delta_2-\langle U\rangle$.  The analysis that follows is closely related to that which was originally done in Ref.~\cite{myprl}.  Figure~\ref{baren} depicts the bare Fock energies of the unmodulated, energy-offset lattice.  In the absence of any modulation, there are no system degeneracies, and the action of the reservoirs evolve an arbitrary system almost completely into the $|2,0\rangle$ state.

\begin{figure}[h!]
\includegraphics[width=.45\textwidth]{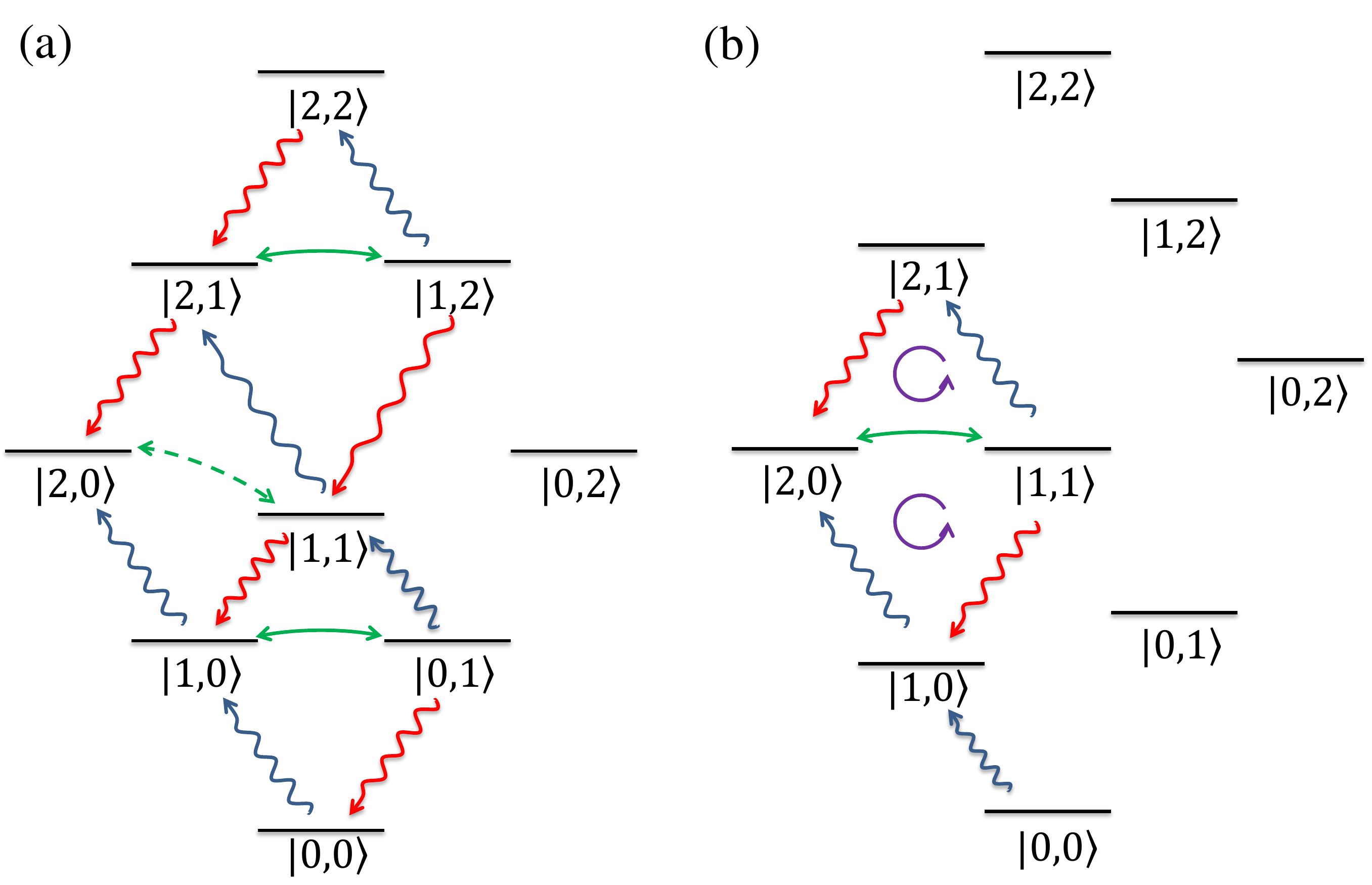}
\caption{Relative energies for $\hat{H}_\text{eff}$ of a two-site lattice with finite $\delta_1$ and $U$ when it is modulated at (a) $\delta_2$ and (b) $\delta_2-\langle U\rangle$.  The blue arrows represent actions of the left reservoir, the red arrows represent actions of the right reservoir, while the green arrows represent intrasystem transitions.  As seen in (a), modulations with $\delta_2$ create an effective resonance between $|1,0\rangle$, and $|0,1\rangle$.  However, the energetic mismatch between $|2,0\rangle$ and $|1,1\rangle$ prohibits effective transport when the reservoir is driving the $|2,0\rangle$ state.  (b)  Alternatively, if the system is modulated with $\delta_2-\langle U\rangle $, the resonance between $|2,0\rangle$ and $|1,1\rangle$ recovers the maximal transport response for this system. }\label{deltarot}
\end{figure}

Figure~\ref{deltarot}(a) displays the energies for the effective Hamiltonian obtained by modulating the system at $\delta_2$.  This establishes a resonance between $|1,0\rangle$ and $|0,1\rangle$, as well as $|2,1\rangle$ and $|1,2\rangle$.  When restricted to the single particle excitation case as in Sec.~\ref{sbs}, this condition optimizes transport since the action of the reservoirs drives population cycles through the lowest four states in the diagram.  However, when $\mu_L$ is set to maintain an occupancy of two particles, the energy mismatch between $|2,0\rangle$ and $|1,1\rangle$ assures that, regardless of the system's initial condition, the action of the reservoirs evolve the vast majority of the system population into the $|2,0\rangle$ state.  Current may leave the system, but only via a second-order, off-resonant transition from $|2,0\rangle\to|0,2\rangle$.  Such a transition leads to the relatively small current gain seen in Fig.~(\ref{3stsw}) when $\alpha\simeq 0.2\times (\delta_2-\langle U\rangle )$. 

Figure~\ref{deltarot}(b) shows the energetic relationship of  Eq.~(\ref{bh2EffGapped}), the effective Hamiltonian associated with modulating the system at $\delta_2-\langle U\rangle$.  This modulation makes the system in question equivalent to the forward-biased atomtronic diode introduced in Ref.~\cite{myprl}, but with $J$ reduced by a factor of $4$.  The resonance between the $|2,0\rangle$ and $|1,1\rangle$ states assures a large transport response via cycles between the $|1,0\rangle$, $|2,0\rangle$, $|1,1\rangle $ and $|2,1\rangle$ states.

Thus appreciable transport enhancement in higher particle manifolds is possible, but care needs to be taken to ensure that the correct resonance conditions are met.

\section{Conclusion}
In this article, we have demonstrated that varying the depth of an optical lattice so that the tunneling barriers experience a sinusoidal modulation can greatly enhance the transport characteristics that are otherwise suppressed by energetic mismatches in the system.  Through a secular approximation, we have shown that this modulation maps a stationary lattice that prohibits transport onto an effective stationary lattice that allows for optimal transport.  We have demonstrated transport recovery  for both single- and double-site occupied lattice sites.  For the case where at most one atom was allowed on any given site, we verified the behavior of our proposed modulation schemes as well as our effective stationary models for a variety of five-site lattices.  We also examined the current response as a function of the modulation intensity and concluded that even very small variations of the lattice depth can generate significant gains in transport.  Finally, we demonstrated how current can be recovered when higher particle manifolds are accessed.  For this case, the conditions for optimizing transport change: the modulation frequencies need to be chosen to create the correct effective resonances across particular particle manifolds in the system.

\section{Acknowledgments}
R. A. Pepino would like to thank M. J. Holland for initially suggesting that modulating tunneling rates might enhance atomic transport in optical lattice systems with energetic mismatches.  He would also like to thank S. T. Rittenhouse, J. J. Kinnunen, D. Meiser and J. Cooper for proof reading this manuscript, as well as for their helpful suggestions.

\end{document}